# Polarized solitons in a cubic-quintic medium

Sergei O. Elyutin*

*Department of Physics, Moscow Engineering Physics Institute, Moscow 115409, Russia*

**ABSTRACT**

The interaction of both scalar and counter-rotating polarized steady state pulses (SSP) is studied numerically for a medium characterized by nonlinear susceptibilities of the third and the fifth order (a cubic-quintic medium with $\chi_3 > 0$, $\chi_5 < 0$). The collision of two plateau-shaped solitons proved to be essentially inelastic, as a number of secondary elliptically polarized solitary waves arise as a result of interaction of steady-state pulses.

## 1. INTRODUCTION

The propagation of short optical pulses in a non-linear dispersive photorefractive medium with the nonlinear properties of the order higher then in Kerr media (i.e. the refractive index can be expanded into series $n = n_0 + n_{Kerr} I + n_4 I^2$) is described by the nonlinear Schrödinger equation, which should be now generalized by including the term responsible for the higher nonlinearity[1-4]. A case of special interest is the competing nonlinearities, e.g. focusing cubic $n_{Kerr} > 0$ and large defocusing quintic $n_4 < 0$ nonlinearity (e.g. *Rb* vapor or nonlinear material P-touline sulfonate[5]). Such competition precludes the propagating pulse collapse[6] and can promote the formation of it as a steady-state pulse.

The existence of the steady-state solutions in the form of both bright and dark solitons and also kinks are well known from different authors[1-3, 7-10], though the found steady-state pulses (SSP) are not strictly speaking solitons. In this paper, the phenomenon of steady–state pulse propagation in a cubic-quintic medium is numerically considered as a problem of pulse interaction and robustness under collision[11, 12].

Two cases are observed. The first is the interaction of scalar pulses where the numerical simulation is able to reveal new features of the temporal behavior of the ultrashort pulses in the dynamical processes of their collisions and co-propagation. For instance, it is demonstrated that disregarding to their energy, the SSPs propagates without the form change, so they can be understood widely as solitons. Two low energy solitons collide elastically, so that the pulses shapes do not alter noticeably after interaction, but the interaction of two plateau-shaped solitons appears to be essentially inelastic causing pulses break down or the forming of a sort of bounded state of two SSPs oscillating in the course of propagation. .

Secondly, we consider the propagation and interaction of counter circularly polarized pulses, which were attempted in[13]. It occurred that the transfer from the interaction of the weak and moder-ate SSPs to a collision of the ultimate pulses brings the novel particularities in conquering effects of Kerr and quintic nonlinearities. The solitary waves arisen as a result of collision of two circularly polarized SSPs further evolve in elliptical and even linear polarized pulses.

*elyutin@mail.ru

## 2. STEADY-STATE SOLUTION OF SCALAR CUBIC-QUINTIC NLS

The wave equation for slowly varying envelope $A(z,t)$ of the optical pulse is equation (1). By omitting the space derivative of second order one can write

$$2i\beta \frac{\partial A}{\partial z} + 2i\beta v_g^{-1} \frac{\partial A}{\partial t} + 2s_2\beta|\sigma|\frac{\partial^2 A}{\partial t^2} + \frac{4\pi\omega^2}{c^2}\left(\chi^{(3)}|A|^2 - s_5\left|\chi^{(5)}\right||A|^4\right)A = 0, \qquad (1)$$

where the high order susceptibilities are the quantities averaged over the transverse modes of some fiber. The dispersion parameter

$$\sigma = \frac{1}{2\beta}\left(\frac{1}{v_g^2} + \beta\frac{\partial^2 \beta}{\partial \omega^2}\right) = \frac{1}{2v_g^2}\left(\frac{1}{\beta} - \frac{\partial v_g}{\partial \omega}\right),$$

$s_2 = \text{sgn}(\sigma)$, $s_5 = +1$ for a self-defocusing non-linearity. The anomalous dispersion corresponds to $s_2 = +1$ in our notations.

In terms of dimensionless independent variables $\zeta, \tau$: $z = L_{cq}\zeta$, $t = t_{cq}\tau + z/v_g$, this equation can be written in a conventional NLS form, providing a CQ-NLS equation:

$$i\frac{\partial e}{\partial \zeta} + \frac{1}{2}\frac{\partial^2 e}{\partial \tau^2} + \left(|e|^2 - |e|^4\right)e = 0, \qquad (2)$$

where the complex field amplitude $e(\zeta,\tau)$ is the dimensionless variable defined as $e = A \cdot A_{cq}^{-1}$. The normalising field $A_{cq}$ is chosen to equalise the cubic and quintic contribution in the combine nonlinearity, i.e. $A_{cq}^2 = \chi^{(3)}\chi^{(5)^{-1}}$. Parameter $L_{cq}$ is the spatial scale of the Kerr self- and cross-modulation process in the field of the $A_{cq}$ strength

$$L_{cq} = \frac{\beta c^2 \chi^{(5)}}{2\pi\omega^2 \chi^{(3)^2}} \approx \frac{n\lambda_0}{4\pi^2 \chi^{(3)} A_{cq}^2}. \qquad (3)$$

The time scale parameter

$$t_{cq} = \left(2\sigma L_{cq}\right)^{1/2}$$

can be interpreted as the duration of the pulse which suffers the dispersion broadening of about the order of magnitude at distance $z = L_{cq}$ propagating in a linear regime, $s_2 = s_5 = +1$.

A steady-state solution of equation (2) was obtained in a number of previous works[1-3, 7-10] in the form of a pulse $e_{st}(\zeta,\tau) = \mathcal{E}(\zeta,\tau)\exp\{i\varphi(\zeta,\tau)\}$. For such representation, equation (2) can be rewritten for real quantities

$$\frac{\partial \mathcal{E}}{\partial \zeta} + \frac{\partial \varphi}{\partial \tau}\frac{\partial \mathcal{E}}{\partial \tau} + \frac{1}{2}\mathcal{E}\frac{\partial^2 \varphi}{\partial \tau^2} = 0, \tag{4.1}$$

$$\frac{1}{2}\left(\frac{\partial^2 \mathcal{E}}{\partial \tau^2} - \mathcal{E}\frac{\partial \varphi}{\partial \tau}\frac{\partial \varphi}{\partial \tau}\right) - \mathcal{E}\frac{\partial \varphi}{\partial \zeta} + \mathcal{E}^3 - \mathcal{E}^5 = 0. \tag{4.2}$$

By assuming, that $\partial \varphi / \partial \tau = \Omega$ and $\partial \varphi / \partial \zeta = K$, from (4.1) one can obtain that $\mathcal{E}_{st}$ depends only on one variable $\eta = \tau + \Omega \zeta$. Then (4.2) transfers into

$$\frac{\partial^2 \mathcal{E}}{\partial \eta^2} - p^2 \mathcal{E} + 2\mathcal{E}^3 - 2\mathcal{E}^5 = 0, \tag{5}$$

where $p^2 = 2K + \Omega^2$ is introduced. The integration of this equation with the boundary conditions corresponding to a solitary wave yields

$$\left(\frac{d\mathcal{E}}{d\eta}\right)^2 = p^2 \mathcal{E}^2 - \mathcal{E}^4 + \frac{2}{3}\mathcal{E}^6. \tag{6}$$

The change of variables $\mathcal{E} = w^{-1/2}$ converts this equation in the following

$$\left(\frac{dw}{d\eta}\right)^2 = 4\left(p^2 w^2 - w + \frac{2}{3}\right) = 4p^2\left[(w-a)^2 - \Delta^2\right], \tag{7}$$

where $a = 1/2p^2$, $\Delta^2 = a^2(1 - 8p^2/3)$. The solution of (7) is

$$w = a + \Delta \cosh[2p(\eta - \eta_0)],$$

where the integration constant $\eta_0$ locates the maximum of a solitary wave.

The final expression for the solution of equation (6) writes in the form:

$$\mathcal{E}^2(\eta) = \frac{2p^2}{1 + (1 - 8p^2/3)^{1/2} \cosh[2p(\eta - \eta_0)]} \tag{8}$$

where $\eta = \tau + \Omega \zeta$, and $\partial \varphi / \partial \tau = \Omega$ и $\partial \varphi / \partial \zeta = K$.

As it is seen from (8) parameter $p$ is bounded by an inequality: $p \leq p_c = 3/8$. That means that the amplitude of a steady-state pulse is also limited

$$\max \mathcal{E}^2(\eta) = \frac{2p^2}{1 + (1 - 8p^2/3)^{1/2}}\bigg|_{p=p_c} = 2p_c^2 = 3/4 \tag{9}$$

or $\max \mathcal{E}(\eta) = (\sqrt{3}/2) \approx 0,866025$.

The energy of a solitary pulse (8)

$$W(p) = 2p_c \ln\frac{p_c + p + \sqrt{p_c^2 - p^2}}{p_c - p + \sqrt{p_c^2 - p^2}} = 4p_c \operatorname{Artanh}\frac{\sqrt{1-b^2}}{1+b}$$

where $b = \sqrt{1 - p^2/p_c^2}$.

The pulse width on half maximum $\Delta\tau_{st}$ is defined from (8) as

$$\Delta\tau_{st} = \frac{1}{p}\text{Arcosh}(Z) = \frac{1}{p}\ln\left(Z + \sqrt{Z^2 - 1}\right), \tag{10}$$

for $Z = 2 + 1/b = 2 + p_c(p_c^2 - p^2)^{-1/2}$.

It should be noted that with the growth of $p$ the energy of the pulse increases monotonically while its width first declines then grows. That means that the function of the pulse width vs energy is not monotonic. Fig.1 depicts $\Delta\tau_{st}$ and $W(p)$ as a functions of $p$.

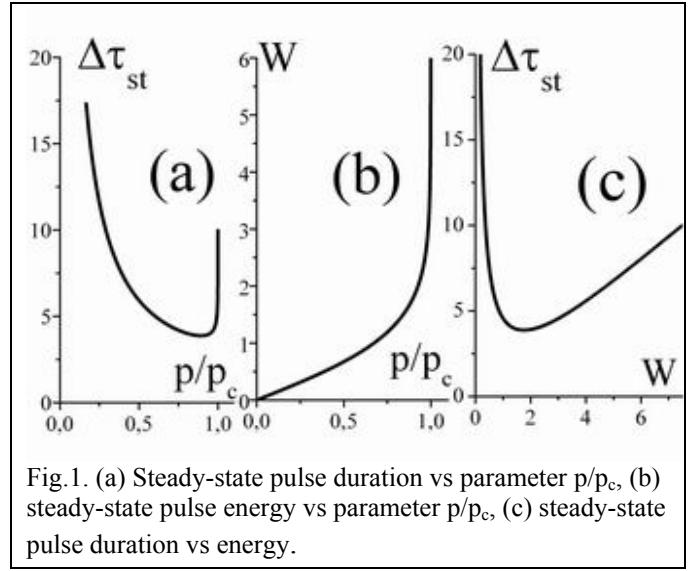

Fig.1. (a) Steady-state pulse duration vs parameter $p/p_c$, (b) steady-state pulse energy vs parameter $p/p_c$, (c) steady-state pulse duration vs energy.

## 3. NUMERICAL ESTIMATES

A large deviation from the Kerr nonlinearity was observed for nonlinear polymers. As it follows from[5] for the organic nonlinear material P-touline sulfonate (PTS) $n_2 = \chi_{isu}^{(3)}(2n)^{-1} = 10^{-18}$ m²/V², $n_4 = |\chi_{isu}^{(5)}|(2n)^{-1} = 2.5\times10^{-34}$ m⁴/V⁴ (note, that 1V=1/300 CGSE). Bearing in mind that $\chi^{(3)} = (4\pi)^{-1}\chi_{isu}^{(3)}$, $\chi^{(5)} = (4\pi)^{-1}\chi_{isu}^{(5)}$, we obtain $\chi^{(3)} = 1.7\cdot10^{-10}$ CGSE (compare with $\chi^{(3)} = 2\cdot10^{-14}$ CGSE for an ordinary glass) and $\chi^{(5)} = 5\cdot10^{-17}$ CGSE. We adopt $\lambda_0 = 2\pi c\omega^{-1} \approx 1.06\mu m$ as the wave length in vacuum, $\beta \approx 2\pi n\lambda_0^{-1}$ as the propagation constant, $D = 4\pi c\sigma\lambda_0^{-2} = 1500$ ps/nm/km $\approx 1.5\cdot10^{-7}$ s·cm⁻². The dispersion parameter $\sigma \approx 10^{-26}$ s²·cm⁻¹ is a typical value for glasses. These values provide the following estimations for the scale parameters of the current problem

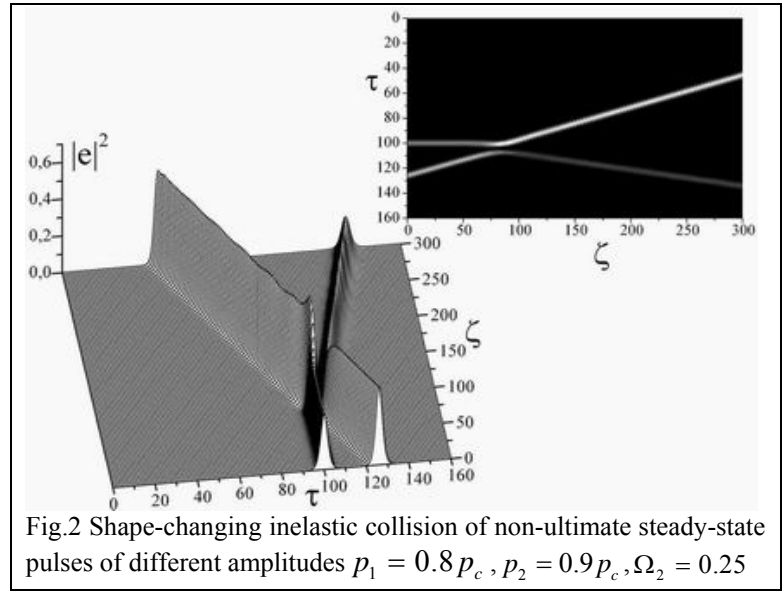

Fig.2 Shape-changing inelastic collision of non-ultimate steady-state pulses of different amplitudes $p_1 = 0.8p_c$, $p_2 = 0.9p_c$, $\Omega_2 = 0.25$

$$A_{cq} = \left(\chi^{(3)}\chi^{(5)-1}\right)^{1/2} \approx 1.8\cdot10^3 \text{ CGSE}, \quad L_{cq} \approx \frac{n\lambda_0}{4\pi^2\chi^{(3)}A_{cq}^2} = \frac{n\lambda_0\chi^{(5)}}{4\pi^2\chi^{(3)^2}} \approx 15 \text{ μm}, \quad t_{cq} = \left(2\sigma L_{cq}\right)^{1/2} \approx 5 \text{ fs}.$$

The magnitude of $\chi^{(5)}$ for the conventional materials such as most of glasses is not reliably measured, so one can only speculate that it is considerably less than it is for PTS.

Let set it about $\chi^{(5)} = 5 \cdot 10^{-19}$ CGSE and $\chi^{(3)} = 2 \cdot 10^{-14}$ CGSE, then $A_{cq} \approx 2 \times 10^2$ CGSE.

$$L_{cq} \approx \frac{n\lambda_0}{4\pi^2 \chi^{(3)} A_{cq}^2} = \frac{n\lambda_0 \chi^{(5)}}{4\pi^2 \chi^{(3)^2}} \approx 30 \text{ m} \quad t_{cq} = \left(2\sigma L_{cq}\right)^{1/2} \approx 8 \text{ ps}$$

### 4. MODELING OF PULSE INTERACTION IN A SCALAR CUBIC-QUINTIC NLS

The first that is interesting to observe in numerical simulation is a collision of two scalar pulses. The colliding pulses suffer a strong perturbation, which could cause their instability.

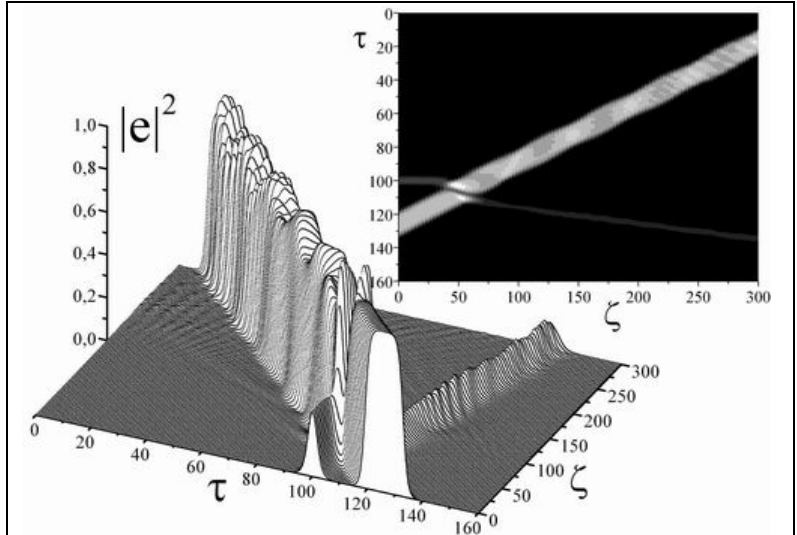

Fig.3. Collision of ultimate and weak steady-state pulses $p_1 = 0.8 p_c$, $p_2 = 0.99999997019 p_c$

The collision of two moderate ($p_1, p_2 < p_c$) pulses demonstrates steadiness accompanying by the amplitude exchange and a slight deviation in the trajectory directions (Fig.2). The whole picture is very similar to the conventional NLS soliton collision that is quite clear as for the weak field the main contribution to the pulse dynamics is made by conquer mechanism of dispersion and Kerr cubic nonlinearity. On the contrary, a collision of two pulses, when at least one is an ultimate plateau-shaped pulse $p_1 \square p_c$ with the peak intensity 3/4 (9), exhibits evident lack of robustness as it is seen in Fig.3. A strike of the low energy pulse, though not very strong as the spot of interaction is small, disbalance the process of competition between the Kerr squeezing and the quintic temporal broadening on the top of the ultimate pulse. The result is the developing of internal oscillation of the time shape

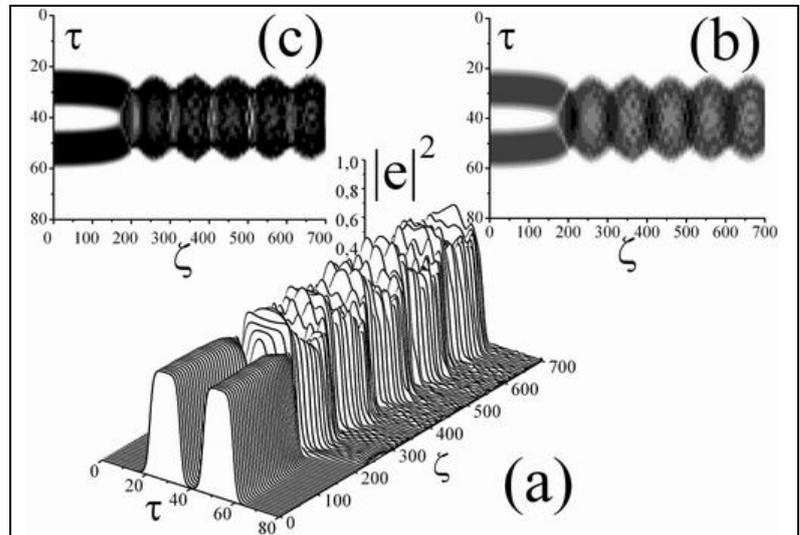

Fig.4 (a) Propagation of two similar closely localized ultimate SSPs for $p_1, p_2 = 0.9999999 p_c$; (b) The grey scale map of the process; (c) The grey scale map of a spatial-temporal distribution of the normalized refraction index $n/n_0$ with light grey up to dark grey corresponds to the transition from unit and higher.

of the ultimate pulse in the course of its further propagation. The radiation of weak quasiharmonic waves from the vicinity of pulse trajectory should be noted as a feature of a non-solition behavior.

The role of a perturbation in a dynamic of ultimate pulses looks even brighter in a numerical experiment when two co-propagating ultimate pulses weakly affect each other at the entrance to a fiber by their wings (Fig.4(a,b)). In the inset (c) to Fig.4 we demonstrate a spatial-temporal evolution of a nonlinear refraction index $n/n_0$ as a result of interplay between dispersion, focusing Kerr nonlinearity and stronger defocusing quintic nonlinearity. That complex reaction of a transparent medium causes an internal oscillations, mentioned in[14], on the background of a stable further propagation of a combined solitary wave at many distance units in fiber. It should be emphasized at this point that the numerical experiment in Fig.4 disputes some conclusions in[3], where the repelling of two co-propagating solitons was predicted.

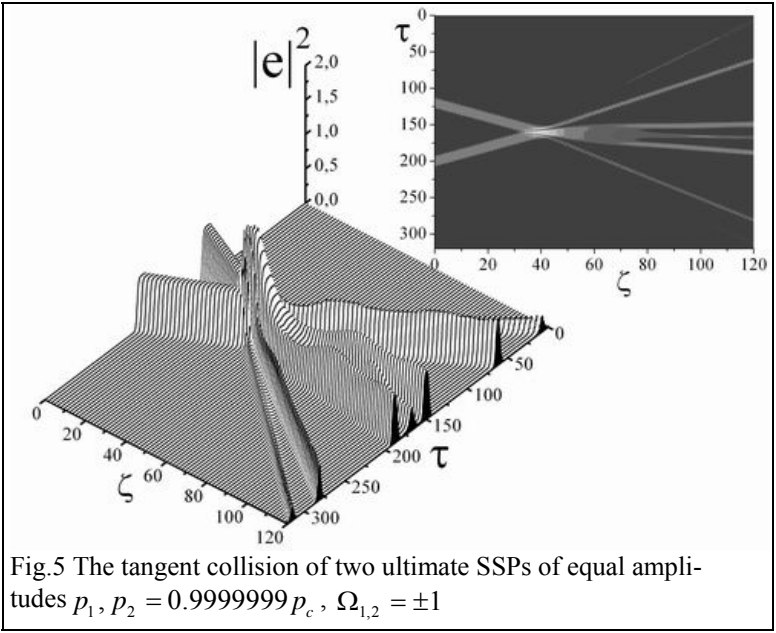

Fig.5 The tangent collision of two ultimate SSPs of equal amplitudes $p_1, p_2 = 0.9999999 p_c$, $\Omega_{1,2} = \pm 1$

The collision of two ultimate pulses (Fig.5) with $\Omega_{1,2} = \pm 1$ due to a strong perturbation of solitary waves (the spot of interaction is relatively large) produces the dramatic result when both pulses collapse giving rise to a family of weak pulses with medium values of $p_i < p_c$ running off the place of collisions with different phase velocities.

Fig.6. depicts the propagation of a pulse with ultimate amplitude but with the pulse width $\delta_{pulse} = 4\delta_{ultimate}$. This pulse can be conditionally considered as containing several SSPs. The grey scale map clearly displayed the existence of a regular periodic motion inside the field of pulse propagation.

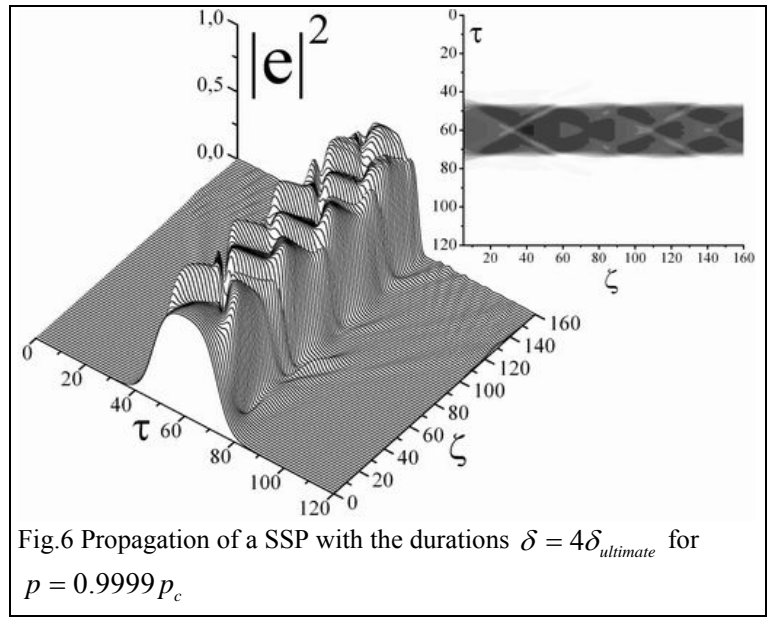

Fig.6 Propagation of a SSP with the durations $\delta = 4\delta_{ultimate}$ for $p = 0.9999 p_c$

## 5. THE VECTOR EXPANDING OF A CUBIC-QUINTIC NLS

Except the obvious generalization of a CQ-NLS, in which a module of a complex vector $\mathbf{e} = (e_1, e_2)$ substitutes the module of a complex function $e$ in (2), one could offer many other versions of a CQ-

NLS. Here we follow[15] in formulation of a set of equation for circularly counter-polarized waves (Vector CQ-NLS):

$$i\frac{\partial e_1}{\partial \zeta} + \frac{1}{2}\frac{\partial^2 e_1}{\partial \tau^2} + \left(|e_1|^2 + \beta|e_2|^2\right)e_1 - \left(|e_1|^4 + 2\alpha|e_1|^2|e_2|^2 + \alpha|e_2|^4\right)e_1 = 0, \quad (11.1)$$

$$i\frac{\partial e_2}{\partial \zeta} + \frac{1}{2}\frac{\partial^2 e_2}{\partial \tau^2} + \left(|e_2|^2 + \beta|e_1|^2\right)e_2 - \left(|e_2|^4 + 2\alpha|e_1|^2|e_2|^2 + \alpha|e_1|^4\right)e_2 = 0. \quad (11.2)$$

with $\alpha=2$, $\beta=3$. Under the condition $e_2 = 0$ the equation for $e_1$ (11.1) transfers into (2).

The Hamiltonian the system comes from as a set of Hamiltonian equations has the form:

$$\mathcal{H} = \mathcal{H}_1 + \mathcal{H}_2 + \mathcal{H}_{int},$$

where $\mathcal{H}_{1,2} = \frac{1}{2}\left|\frac{\partial e_{1,2}}{\partial \tau}\right|^2 - \frac{1}{2}|e_{1,2}|^4 + \frac{1}{3}|e_{1,2}|^6$, $\mathcal{H}_{int} = -\beta|e_1|^2|e_2|^2 + \alpha\left(|e_1|^4|e_2|^2 + |e_1|^2|e_2|^4\right)$.

Let again $e_{1,2}(\zeta,\tau) = \mathcal{E}_{1,2}(\zeta,\tau)\exp\{i\varphi_{1,2}(\zeta,\tau)\}$, and then the system of equations (11) can be rewritten for the real variables:

$$\frac{\partial \mathcal{E}_1}{\partial \zeta} + \frac{\partial \varphi_1}{\partial \tau}\frac{\partial \mathcal{E}_1}{\partial \tau} + \frac{1}{2}\mathcal{E}_1\frac{\partial^2 \varphi_1}{\partial \tau^2} = 0, \qquad \frac{\partial \mathcal{E}_2}{\partial \zeta} + \frac{\partial \varphi_2}{\partial \tau}\frac{\partial \mathcal{E}_2}{\partial \tau} + \frac{1}{2}\mathcal{E}_2\frac{\partial^2 \varphi_2}{\partial \tau^2} = 0 \quad (12.1)$$

$$\frac{1}{2}\left(\frac{\partial^2 \mathcal{E}_1}{\partial \tau^2} - \mathcal{E}_1\frac{\partial \varphi_1}{\partial \tau}\frac{\partial \varphi_1}{\partial \tau}\right) - \mathcal{E}_1\frac{\partial \varphi_1}{\partial \zeta} + \left(\mathcal{E}_1^2 + \beta\mathcal{E}_2^2\right)\mathcal{E}_1 - \left(\mathcal{E}_1^4 + 2\alpha\mathcal{E}_1^2\mathcal{E}_2^2 + \alpha\mathcal{E}_2^4\right)\mathcal{E}_1 = 0. \quad (12.2)$$

$$\frac{1}{2}\left(\frac{\partial^2 \mathcal{E}_2}{\partial \tau^2} - \mathcal{E}_2\frac{\partial \varphi_2}{\partial \tau}\frac{\partial \varphi_2}{\partial \tau}\right) - \mathcal{E}_2\frac{\partial \varphi_2}{\partial \zeta} + \left(\mathcal{E}_2^2 + \beta\mathcal{E}_1^2\right)\mathcal{E}_2 - \left(\mathcal{E}_2^4 + 2\alpha\mathcal{E}_1^2\mathcal{E}_2^2 + \alpha\mathcal{E}_1^4\right)\mathcal{E}_2 = 0. \quad (12.3)$$

Concerning a steady-state solution, it is natural to suggest that both polarization components of the pulse propagate with the same group velocity, i.e. the relations $\partial\varphi_1/\partial\tau = \partial\varphi_2/\partial\tau = -\Omega$ must hold. Both real envelopes $\mathcal{E}_{1,2}(\zeta,\tau)$ then depend on one variable $\eta = \tau + \Omega\zeta$. For the sake of simplicity, one can additionally impose $\partial\varphi_1/\partial\zeta = \partial\varphi_2/\partial\zeta = K$.

The system of equations for the real envelopes (12.2) and (12.3) can be written in the following from:

$$\frac{d^2\mathcal{E}_1}{d\eta^2} = p_1^2\mathcal{E}_1 - 2\left(\mathcal{E}_1^2 + \beta\mathcal{E}_2^2\right)\mathcal{E}_1 + 2\left(\mathcal{E}_1^4 + 2\alpha\mathcal{E}_1^2\mathcal{E}_2^2 + \alpha\mathcal{E}_2^4\right)\mathcal{E}_1, \quad (13.1)$$

$$\frac{d^2\mathcal{E}_2}{d\eta^2} = p_2^2\mathcal{E}_2 - 2\left(\mathcal{E}_2^2 + \beta\mathcal{E}_1^2\right)\mathcal{E}_2 + 2\left(\mathcal{E}_2^4 + 2\alpha\mathcal{E}_1^2\mathcal{E}_2^2 + \alpha\mathcal{E}_1^4\right)\mathcal{E}_2, \quad (13.2)$$

where parameters $p_{1,2}^2 = 2K_{1,2} + \Omega^2$ are introduced.

Symmetrical solution $\mathcal{E}_1 = \mathcal{E}_2 = \mathcal{E}$ and $p_1 = p_2 = p$ is the easiest case to consider. Equations (13) can be now once integrated to produce an equation

$$\left(\frac{d\mathcal{E}}{d\eta}\right)^2 = p^2 \mathcal{E}^2 - (1+\beta)\mathcal{E}^4 + \frac{2\kappa}{3}\mathcal{E}^6, \tag{14}$$

with $\kappa = 1 + 3\alpha$. The solution of this equation can be found in the manner similar to CQ-NLS (2). Its final view is as following:

$$\mathcal{E}^2(\eta) = \frac{2p^2(1+\beta)^{-1}}{1 + (1 - p^2/p_c^2)^{1/2} \cosh[2p(\eta - \eta_0)]}, \tag{15}$$

where parameter $p \leq p_c$, the critical value of $p$ is set by the formula:

$$p_c^2 = 3/8\,(1+\beta)^2(1+3\alpha)^{-1}.$$

By defining the energy of a solitary bi-component pulse as

$$W(p) = \int_{-\infty}^{+\infty} (|e_1|^2 + |e_2|^2)\,d\eta,$$

for the obtained symmetric solution one can find out that

$$W(p) = \frac{4p}{1+\beta} \int_0^{+\infty} \frac{dx}{1 + b\cosh x},$$

where $b = \sqrt{1 - p^2/p_c^2}$. The final expression writes in the form

$$W(p) = \frac{4p_c}{1+\beta} \ln \frac{p_c + p + \sqrt{p_c^2 - p^2}}{p_c - p + \sqrt{p_c^2 - p^2}}. \tag{16}$$

It follows from (15) that the width of a SSP at the half-maximum $\Delta \tau_s$ is

$$\Delta \tau_s = \frac{1}{p}\mathrm{Arcosh}(Z) = \frac{1}{p}\ln\left(Z + \sqrt{Z^2 - 1}\right) \tag{17}$$

for $Z = 2 + 1/b = 2 + p_c(p_c^2 - p^2)^{-1/2}$. This expression is the analogue of formula (10) for a scalar pulse. The difference is only in the value of parameter $p_c$.

## 6. NUMERICAL SIMULATION OF PULSE INTERACTION IN A VECTOR CUBIC-QUINTIC NLS

The numerical study demonstrates that disregarding to their energy, the circularly polarized SSP being launched in each polarization mode individually at the absence of another mode propagates in a soliton fashion without the form changing.

The VCQ-NLS equations (11) contain the inter mode interaction in their cubic and quintic nonlinear terms.

The simulation showed that two low energy solitons ($p_1, p_2 < p_c$), having opposite phase $\Omega_{1,2} = \pm 1$ and spreading each in its own space feel the presence of the counterpart pulse at the field of collision, but the pulses do not alter noticeably after interaction.

As the pulse parameters $p_1$, $p_2$ tend to a critical value $p_c$, the interaction of two crossing moderate solitons of different polarization reveal inelasticity as each input SSP break down into a number of secondary solitary waves runoff the place of intersection with different phase velocities (Fig.7). The observed picture demonstrates a new scenario of an inelastic SSPs interaction, when the cross action of Kerr and quintic nonlinearities triggers off the process of steady state pulses decay into a series of polarised vector pulses with the different degree of ellipticity. Fig.8 (a,b) depicts the collision of ultimate plateau-shaped which proved to be essentially inelastic. There was only one vector pulse from the batch of pulses born in nonlinear interaction, which corresponds to a soliton (pointed by arrows in the insets). This pulse represent linear polarised wave. The others are the solitary waves of elliptical polarization. Though a stability of the elliptically polarized waves in their further propagation is not proved analytically, the numerical simulation in the depth of a medium confirms their steady-state nature.

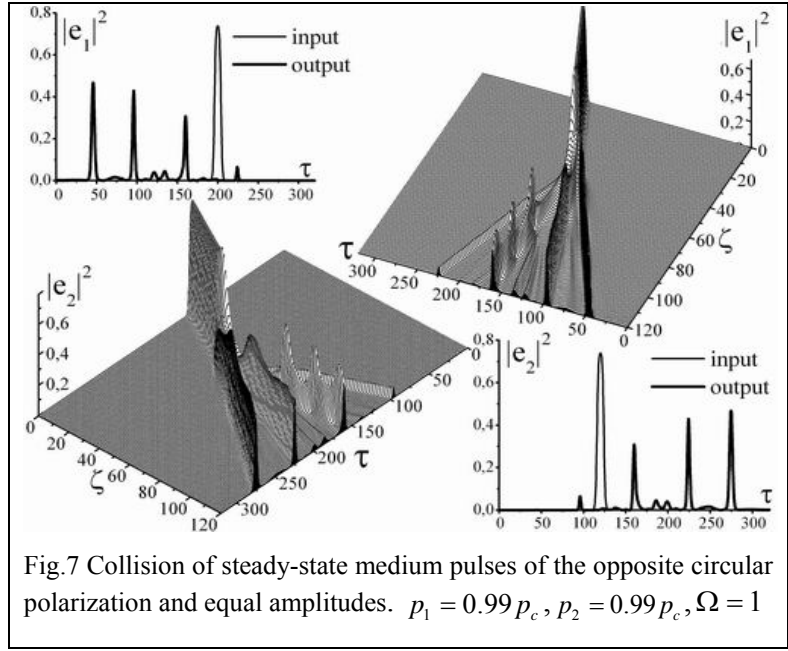

Fig.7 Collision of steady-state medium pulses of the opposite circular polarization and equal amplitudes. $p_1 = 0.99 p_c$, $p_2 = 0.99 p_c$, $\Omega = 1$

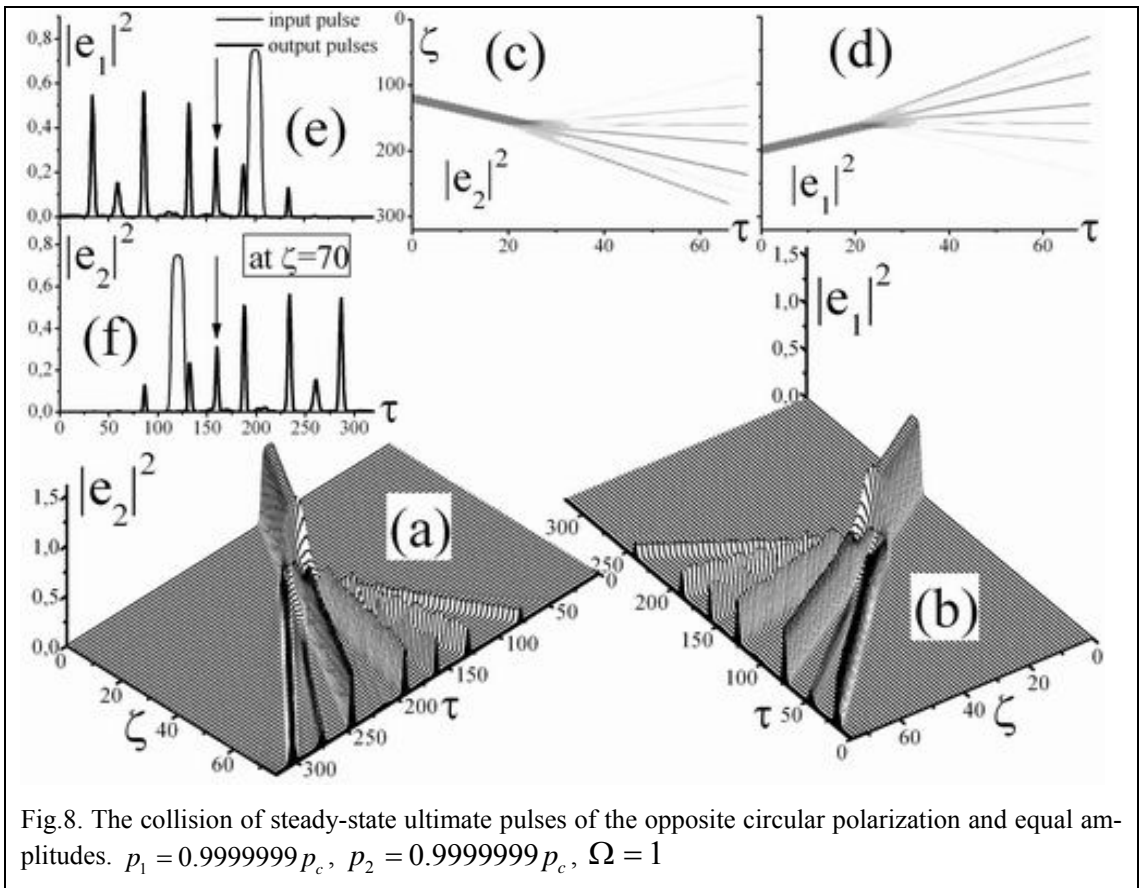

Fig.8. The collision of steady-state ultimate pulses of the opposite circular polarization and equal amplitudes. $p_1 = 0.9999999 p_c$, $p_2 = 0.9999999 p_c$, $\Omega = 1$

The co-linear propagation of two polarized SSP's separated by a short time interval presented another type of interaction (Fig.9 (a,b)). The numerical modelling fixes a split of an input SSP into a family of solitary waves, all but one in the centre moving apart the initial trajectory with different velocities. The temporal profile of the output field shows the bunch of solitary pulses including the linear polarized soliton. (Fig.9 (e,f)). The velocity of the linear polarized soliton (pointed by arrows in Fig.9(e,f)) differ maximum from the velocities of elliptically polarized secondary SSPs. Fig.9(c,d) depicts the grey scale map of the interaction dynamics for the pulses of the counter-rotating polarization. The essence of the observed picture is that the solitary wave on its trail pro-duces a local nonlinear variation of the refractive index in the vicinity of the counterpart SSP, thus dragging a packet of weaker pulses out from the main pulse body. These secondary pulses tend to convert in a linear polarised vector solitons that will probably happen deeper in the medium.

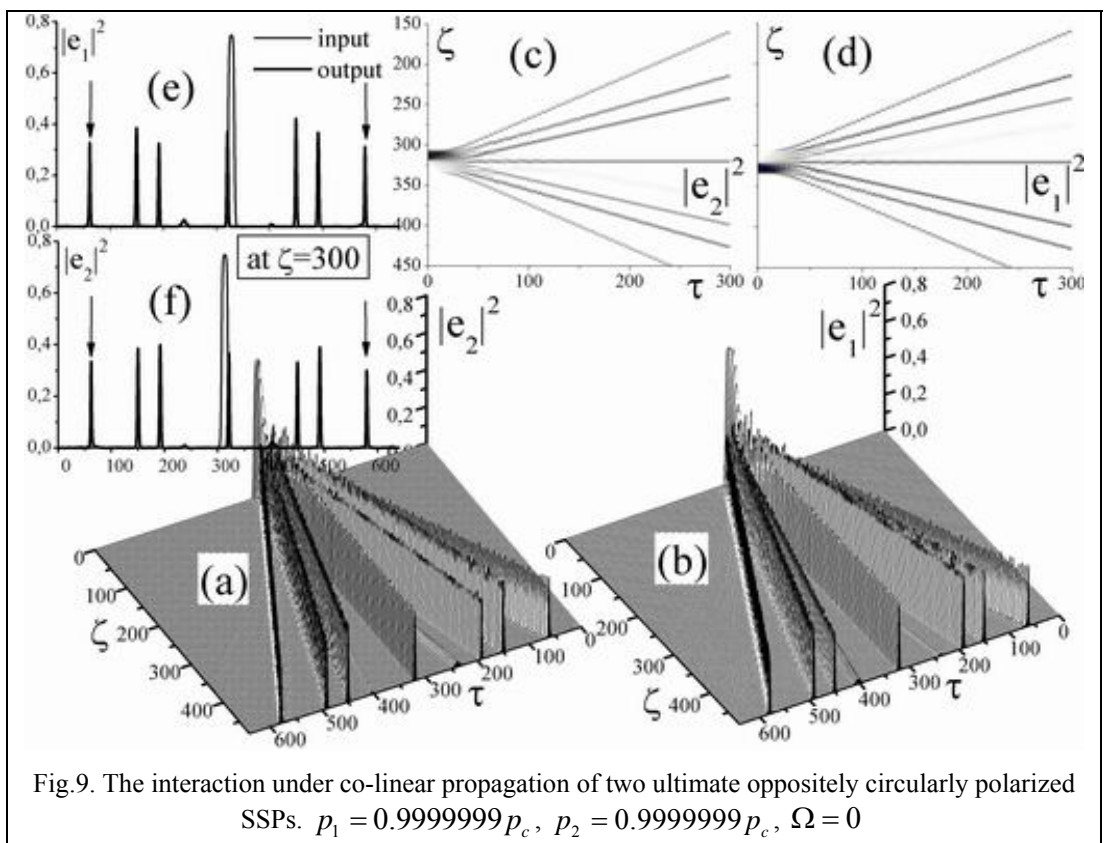

Fig.9. The interaction under co-linear propagation of two ultimate oppositely circularly polarized SSPs. $p_1 = 0.9999999 p_c$, $p_2 = 0.9999999 p_c$, $\Omega = 0$

## ACKNOWLEDGMENTS

The author is grateful to A.I. Maimistov and A.S. Desyatnikov for fruitful discussion. This work is supported by the RFBR grant 06-02-16406

# REFERENCES

1. C. De Angelis, "Self-Trapped Propagation in the Nonlinear Cubic- Quantic Schrodinger Equation: A Variational Approach". *IEEE J. Quant. Electr*. **QE-30**, pp. 818-821, 1994

2. N. Akhmediev, V.V. Afanasjev, "Novel arbitrary-amplitude soliton solutions of the cubiq-quantic-complex Ginzburg-Landau equation", *Phys. Rev. Lett*. **75**, pp 2320 -2323, 1995

3. A.V. Buryak, N.N. Akhmediev, "Internal Friction between Solitons in Near-Integrable Systems", *Phys. Rev*. **E50**, pp. 3126 -3133, 1994

4. B.S. Azimov, M.M. Sagatov, A.P. Sukhorukov, "Formation and propagation of steady-state laser pulses in media with a joint action of nonlinearities of the third and the fifth order", *Kvantovaja Electronica,* **18**, pp. 104-106,1991

5. B. Lawrence, W.E. Torrullas, M. Cha, M.L. Sundheimer, G.I. Stegeman, J. Meth, S. Etamad, G. Baker, "Identification and role of two-photon excited states in a π-conjucated poltymer", *Phys. Rev. Lett.*, **73**, pp. 597-600, 1994

6. B.S. Azimov, M.M.Sagatov, "On effect of explosive compression of laser pulses", *Vestnik MGU*, **32**, pp. 96-97, 1991

7. A.I. Maimistov, A.M. Basharov, *Nonlinear Optical Waves*, Kluwer Academic. Dordrecht, 1999

8. D.E. Pelinovsky, Y.S. Kivshar, V.V. Afanasjev, "Instability-induced dynamics of dark solitons," *Phys. Rev. E*, **54**, pp. 2015-2032, 1996

9. D. Pushkarov, S. Tanev, "Bright and dark solitary wave propagation and bistability in the anomalous dispersion region of optical waveguides with third- and fifth-order nonlinearities", *Opt. Commun.*, **124**, pp. 354-364, 1996

10. D. Artigas, L. Torner, J.P. Torres, N.N. Akhmediev, " Assymetrical splitting of higher-order solitons induced by quintic nonlinearity", *Opt. Commun.*, **143**, pp. 322-328, 1997

11. W. S. Kim, H.-T. Moon "Soliton-kink interaction in a generalized nonlinear Schrodinger system", *Phys. Lett.*, **A266**, pp. 364-369, 2000

12. Y. Chen ,"Dark solitons in weakly saturable nonlinear media", *Phys. Rev.*, **E55**, pp. 1221-1224, 1997

13. R. Radhakrishnan, A. Kundu, and M. Lakshmanan, "Coupled nonlinear Schrodinger equations with cubic-quintic nonlinearity: Integrability and soliton interaction in non-Kerr media", *Phys. Rev.*, **E60**, pp. 3314-3323, 1999

14. G.D. Peng, B.A. Malomed, P.L. Chu, "Soliton collision in a model of dual-core nonlinear optical fiber", *Phys. Scripta*, **58**, pp. 149-158, 1998

15. A. Maimistov, B. Malomed, A. Desyatnikov, "A potential of incoherent attraction between multidimensional solitons," *Phys. Lett*. **A254**, pp. 179-184, 1999